\documentclass[10pt,conference]{IEEEtran}
\usepackage{cite}
\usepackage{amsmath,amssymb,amsfonts}
\usepackage{algorithmic}
\usepackage{graphicx}
\usepackage{textcomp}
\usepackage{xcolor}
\usepackage[hyphens]{url}
\usepackage{fancyhdr}
\usepackage{hyperref}
\usepackage{array}
\usepackage{subcaption}
\usepackage{makecell}
\usepackage{bm}

\title{ProactivePIM : Accelerating Weight-Sharing Embedding Layer with PIM for Scalable Recommendation System}

\newcommand\hpcaaffiliation{}
\newcommand\hpcaemail{}

\author{
    \IEEEauthorblockN{Youngsuk Kim$\dagger$, Junghwan Lim$\dagger$, Hyuk Jae Lee$\dagger$, and Chae Eun Rhee$\ddagger$}
      \IEEEauthorblockA{
        \hpcaaffiliation{Seoul National University$\dagger$, Hanyang University$\ddagger$} \\
        \hpcaemail{\{youngsuk95, jhlim, hjlee\}@capp.snu.ac.kr, crhee@hanyang.ac.kr}
      }
}

\fancypagestyle{camerareadyfirstpage}{%
  \fancyhead{}
  
  \fancyhead[C]{
    \ifdefined\aeopen
    \parbox[][12mm][t]{13.5cm}{\hpcayear{} IEEE International Symposium on High-Performance Computer Architecture (HPCA)}    
    \else
      \ifdefined\aereviewed
      \parbox[][12mm][t]{13.5cm}{\hpcayear{} IEEE International Symposium on High-Performance Computer Architecture (HPCA)}
      \else
      \ifdefined\aereproduced
      \parbox[][12mm][t]{13.5cm}{\hpcayear{} IEEE International Symposium on High-Performance Computer Architecture (HPCA)}
      \else
      \parbox[][0mm][t]{13.5cm}{\hpcayear{} IEEE International Symposium on High-Performance Computer Architecture (HPCA)}
    \fi 
    \fi 
    \fi 
    \ifdefined\aeopen 
      \includegraphics[width=12mm,height=12mm]{ae-badges/open-research-objects.pdf}
    \fi 
    \ifdefined\aereviewed
      \includegraphics[width=12mm,height=12mm]{ae-badges/research-objects-reviewed.pdf}
    \fi 
    \ifdefined\aereproduced
      \includegraphics[width=12mm,height=12mm]{ae-badges/results-reproduced.pdf}
    \fi
  }
  \fancyfoot[C]{}
}
\begin{document}
\maketitle

\ifdefined\hpcacameraready 
  \thispagestyle{camerareadyfirstpage}
  \pagestyle{empty}
\else
  \thispagestyle{plain}
  \pagestyle{plain}
\fi

\newcommand{\hpcaheight}{0mm}
\ifdefined\eaopen
\renewcommand{\hpcaheight}{12mm}
\fi

\begin{abstract}
Although deep learning-based personalized recommendation systems provide qualified recommendations, they strain data center resources. The main bottleneck is the embedding layer, which is highly memory-intensive due to its sparse, irregular access patterns to embeddings. Recent near-memory processing (NMP) and processing-in-memory (PIM) architectures have addressed these issues by exploiting parallelism within memory. However, as model sizes increase year by year and can exceed server capacity, inference on single-node servers becomes challenging, necessitating the integration of model compression. Various algorithms have been proposed for model size reduction, but they come at the cost of increased memory access and CPU–PIM communication.

We present ProactivePIM, a PIM system tailored for weight-sharing algorithms, a family of compression methods that decompose an embedding table into compact subtables, such as QR-trick and TT-Rec. Our analysis shows that embedding layer execution with weight-sharing algorithms increases memory access and incurs CPU–PIM communication. We also find that these algorithms exhibit unique data locality characteristics, which we name intra-GnR locality. ProactivePIM accelerates weight-sharing algorithms by utilizing a heterogeneous HBM-DIMM memory architecture with integration of a two-level PIM system of base-die PIM (bd-PIM) and bank-group PIM (bg-PIM) inside the HBM. To gain further speedup, ProactivePIM prefetches embeddings with high intra-GnR locality into an SRAM cache within bg-PIM and eliminates the CPU-PIM communication through duplication of target subtables across bank groups. With additional optimization techniques, our design effectively accelerates weight-sharing algorithms, achieving $2.22\times$ and $2.15\times$ speedup in QR-trick and TT-Rec, respectively, compared to the baseline architecture.




\end{abstract}


\maketitle

\section{Introduction}
\label{sec:introduction}

Nowadays, a deep learning based personalized recommendation system is a pivotal technology in various companies, including Meta \cite{naumov2019deep}, YouTube \cite{covington2016deep}, and Netflix \cite{steck2021deep}. To generate a qualified prediction, recommendation systems utilize user information (dense features) and interacted items  (sparse features) as inputs, where sparse features are fed into the embedding layer. The embedding layer transforms items into embeddings and executes a gather-and-reduce (GnR) operation on embeddings to capture the user’s preference. Unfortunately, these recommendation systems account for a significant computational workload in data centers \cite{gupta2020architectural}, with the embedding layer being the primary operation that governs the overall production performance. The GnR operation exhibits sparse and irregular patterns, making the embedding layer highly memory-intensive. 

\begin{figure}[htbp]
\centering
\includegraphics[width=0.78\columnwidth]{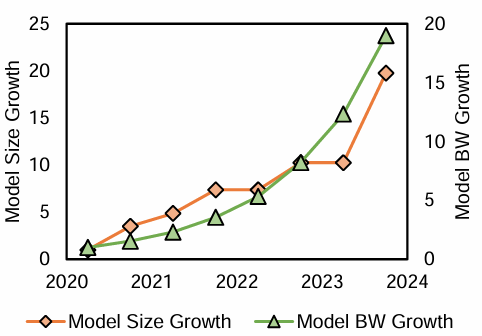}
\caption{Model size and bandwidth growth. (Reproduced from MTIA \cite{firoozshahian2023mtia} and RecShard \cite{sethi2022recshard})
}
\label{fig:dlrm_growth}
\end{figure}

Figure \ref{fig:dlrm_growth} depicts the current trend of a representative recommendation model used in production, namely deep learning recommendation model (DLRM). The memory-intensive nature of the embedding layer has been growing continuously, placing constraints on the data center infrastructure \cite{zhao2022understanding}. Recent studies leveraged near-memory processing (NMP) and processing-in-memory (PIM) to accelerate AI models by exploiting memory parallelism \cite{he2020newton, ke2020recnmp, kwon2019tensordimm, lee2021hardware, park2024attacc}. For recommendation systems, TensorDIMM \cite{kwon2019tensordimm}, RecNMP \cite{ke2020recnmp}, and TRiM \cite{park2021trim} successfully accelerated GnR of the embedding layer via NMP/PIM systems. These approaches effectively exploit the internal memory bandwidth, which is further amplified by the number of ranks in a memory channel. In addition, SPACE \cite{kal2021space} proposed an HBM equipped with an NMP to overcome the inefficient power consumption of DIMM-based NMP/PIM architectures.

On the other hand, as shown in Figure \ref{fig:dlrm_growth}, the model size of DLRM has grown 16 times from 2020 to 2024. The model’s growth is attributed to the incorporation of more item, which generally achieves better model quality \cite{mudigere2022software, sethi2022recshard}. Serving inferences with a multi-node server or introducing solid-state drives (SSDs) could be solutions, but they produce significant synchronization costs, vulnerability to failures, and a negative impact on execution time \cite{ardestani2022supporting, zhao2019aibox}. To mitigate model size growth, the hashing trick \cite{weinberger2009feature} is utilized to fit the model into the memory capacity budget at the cost degradation in accuracy. Weight-sharing algorithms \cite{shi2020compositional, yin2021tt} that transforms the original embedding table into multiple smaller subtables is robust to this drawback and significantly reduce the model size at the same time. There are other algorithmic methods \cite{ginart2021mixed, lyu2022optembed}, but the weight-sharing algorithm is superior in terms of the degree of compression. Nevertheless, these methods amplify off-chip memory access and are still memory-intensive. Direct application of NMP/PIM architectures designed for recommendation system inference to the embedding layer with weight-sharing algorithms is an inefficient approach. The memory access is more than double that of the original embedding layer, increasing the embedding layer latency as well as its execution time with NMP/PIM. Furthermore, the multiplication used in these algorithms requires operands from subtables to reside in the same NMP/PIM unit, unlike the simpler GnR. This necessitates CPU-PIM communication, which causes a severe throughput overhead. 

We propose ProactivePIM, a PIM system designed to address embedding layer with weight-sharing algorithms. The ProactivePIM is built on a heterogeneous HBM-DIMM architecture, where a two-level PIM system is placed within the HBM: base-die PIM (bd-PIM) and bank-group PIM (bg-PIM). \textcolor{black}{This hierarchy supports current and future DLRM inference, given the yearly increase in embedding layer memory intensity.} We further analyze weight-sharing algorithms and identify two key insights: 1) embeddings of subtables exhibit intra-GnR locality, which can be reused multiple times within and beyond a single GnR, and 2) the smallest subtables are compact enough to fit within a bank group. To leverage these insights, we integrate an SRAM cache inside bg-PIM to prefetch embeddings with high intra-GnR locality before GnR execution, resulting in reduction in memory access. Our prefetch scheme is designed to benefit from intra-GnR locality while minimizing cache area overhead. To eliminate the CPU-PIM communication overhead, we propose a subtable mapping strategy that duplicates selected subtables across bank groups. Moreover, we optimize TT-Rec inference by reordering the computation to two-stage skinny general matrix-matrix multiplication (GEMM), thereby reducing the redundant memory access. \textcolor{black}{In this way, our method enables scalability via model compression support.} ProactivePIM can also be applied to memory-intensive workloads with skewed data access patterns. This includes applications such as PageRank \cite{gonzalez2012powergraph}, skinny GEMM \cite{cho2021accelerating} and the original DLRM, which take advantage of its static cache and 2-level PIM architecture.  The key contributions of our work are as follows:
\begin{itemize}
\item We propose ProactivePIM, a PIM system for accelerating weight-sharing algorithms. To the best of our knowledge, our study is the first PIM system to address the model size reduction of personalized recommendation systems.
\item We conducted a detailed analysis of weight-sharing algorithms, identifying their drawbacks and unique characteristics, including intra-GnR locality and subtable size imbalance. 
\item We provide an efficient prefetching scheme and a subtable mapping strategy that leverages the characteristics of weight-sharing algorithms. With these methods, ProactivePIM eliminates CPU-PIM communication overhead and achieves effective PIM speedup with minimal extra PIM area overhead. 

\end{itemize}


\section{Background} \label{background}
\subsection{Personalized Recommendation System}

\begin{figure}[htbp]
\centering
\includegraphics[width=0.8\columnwidth]{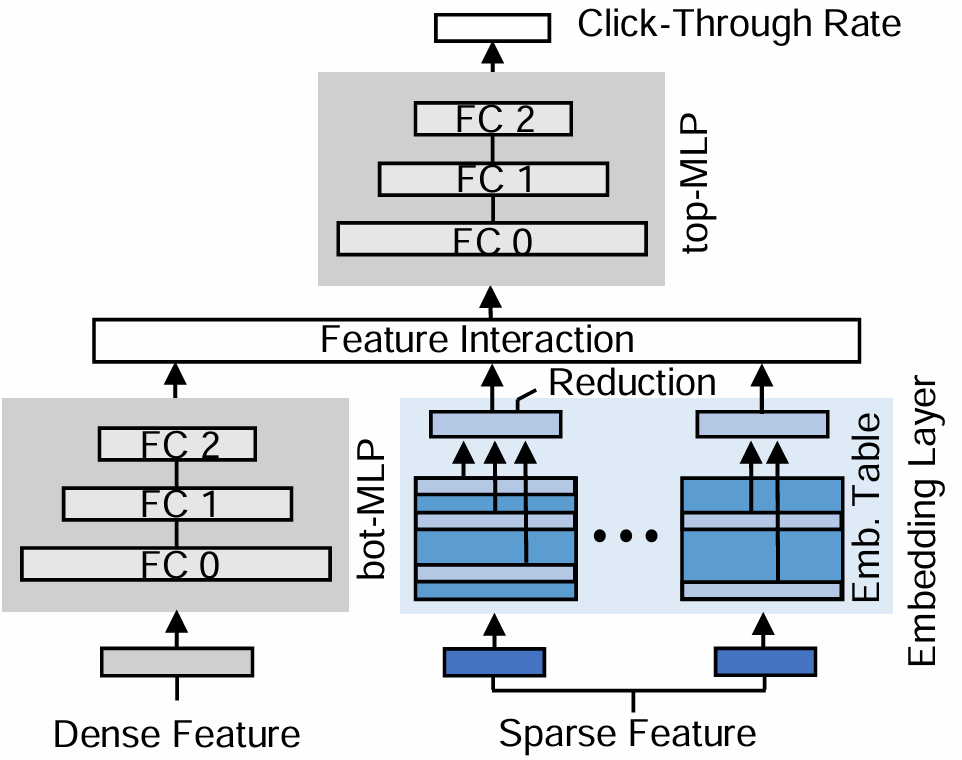}
\caption{DLRM architecture}
\label{fig:dlrm_arch}
\end{figure}

\textbf{Recommendation System Overview.} As shown in Figure \ref{fig:dlrm_arch}, DLRM processes dense features (user information) and sparse features (item interaction history) to predict user preferences. DLRM, a representative recommendation model, comprises three main components: a bottom multi-layer perceptron (MLP), an embedding layer, and a top MLP. The bottom MLP handles dense features, whereas the embedding layer leverages sparse features to extract item embeddings from the embedding table. These outputs are combined through a feature interaction layer, and the result is passed to the top MLP that produces the final click-through-rate (CTR) prediction.

\textbf{Challenges in Recommendation Systems.} Industry-scale DLRM implementations reach sizes of up to tens of terabytes, sometimes surpassing the maximum memory capacity budget of a single-node inference server. Given that the model quality is closely tied to the size of the embedding tables \cite{ardestani2022supporting, zhao2022understanding}, its scale is growing year by year \cite{ardestani2022supporting, firoozshahian2023mtia, sethi2022recshard}. The practical utilization of such large models is constrained by capacity limitations in the server. Conventional hashing methods such as hashing-trick \cite{weinberger2009feature} are employed to fit embedding tables within available memory, but these approaches offer limited effectiveness, as they often incur substantial accuracy degradation. In addition, the memory-intensive nature of the embedding layer poses a further challenge by reducing inference throughput. The sparse feature exhibits irregular and sparse memory access, making it challenging for the embedding layer to benefit from on-chip memory. Combined with the large embedding table size that exceeds the cache capacity, the embedding layer requires frequent access to the off-chip memory system. As the execution of the embedding layer has a longer latency than the bottom MLP \cite{ke2020recnmp, kal2021space}, the memory-bound nature of the embedding layer degrades the inference throughput.


\subsection{NMP/PIM for Recommendation System}
Recent studies have integrated NMP/PIM units into DIMMs to accelerate memory-intensive GnR operations. TensorDIMM \cite{kwon2019tensordimm} and RecNMP \cite{ke2020recnmp} exploited rank-level NMP, whereas TRiM \cite{park2021trim} leveraged a bankgroup-level PIM. These units perform partial GnRs parallel to the data allocated within the memory node. For embedding partitioning, either vertical partitioning (VP) or horizontal partitioning (HP) was utilized. VP divides an embedding across multiple memory nodes equipped with a PIM, whereas HP stores the embedding within a single memory node. The VP is robust to the load imbalance but requires concurrent row activations across memory nodes, resulting in significant power inefficiency. HP suffers from load imbalance owing to irregular access patterns of the embedding layer, but it offers an advantage in terms of power consumption compared to VP. Therefore, previous methods have generally adopted HP.

In pursuit of further speedup, the aforementioned designs leveraged the long-tail distribution of embeddings \cite{kal2021space, ke2020recnmp} with a dynamic cache within the NMP or a load-balancing technique \cite{ke2020recnmp, park2021trim}. To address the challenge of low throughput scalability of DIMM, SPACE utilizes a heterogeneous memory architecture of HBM and DIMM \cite{kal2021space}, using HBM as a static DRAM cache of DIMM to enhance the throughput and power efficiency of the system.

\subsection{Weight-sharing Algorithms for Embedding Table Reduction} \label{weight-sharing-intro}

To mitigate the accuracy degradation of the hashing-trick \cite{weinberger2009feature}, various weight-sharing algorithms have been proposed to address this drawback and achieve better compression, with representative methods including QR-trick \cite{shi2020compositional} and TT-Rec \cite{yin2021tt}. These algorithms reduce the embedding table size by decomposing it into several \textit{subtables} consisting of \textit{subembeddings}. The original embedding must be reconstructed before the GnR operations, which we refer to as collect-and-reconstruction (CnR). CnR typically involves element-wise vector multiplication or GEMV.

\begin{figure}[htbp]
\centering
\includegraphics[width=\columnwidth]{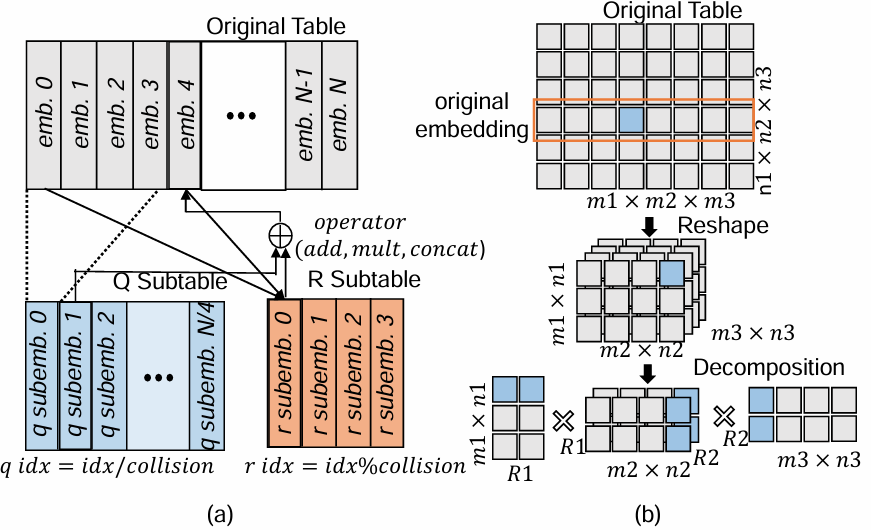}
\caption{Weight-sharing algorithms. (a) QR-trick with hash collision set to 4 (b) TT-Rec with total TT cores set to 3.}
\label{fig:dlrm_intro}
\end{figure}

\textbf{QR-trick.} QR-trick employs a double hashing process to decompose each embedding table into two distinct subtables, where the quotient and remainder operations are used as hash functions. For simplicity, we refer to the subtable generated by the quotient operation as \textit{Q subtable} and that of the remainder operation as \textit{R subtable}. An overview is illustrated in Figure \ref{fig:dlrm_intro} (a). To generate a subtable, each hash function obtains an item index and hash collision as inputs, and maps multiple embedding vectors with the same output to the same subembedding. The Q subtable maps embeddings with identical quotient values to a single subembedding, whereas the R subtable maps embeddings with identical remainder values. QR-trick achieves a compression ratio approximately equal to the inverse of the hash collision. Hash collision value typically range from 1 to 60.

\textbf{TT-Rec. } As shown in Figure \ref{fig:dlrm_intro} (b), TT-Rec compresses each embedding table by decomposing it into several tensor-train cores (TT-cores). Hereafter, we refer to the TT-cores as subtables. The embedding table $T \in \mathbb{R}^{M \times N}$ is reshaped by factorizing $M$ (embedding dimension) and $N$ (total entry of an embedding table) into $M=m1\times m2\times \cdots \times md$ and $N=n1\times n2\times \cdots \times nd$, resulting in a new tensor with a shape of $T \in \mathbb{R}^{(m1 \times n1) \times (m2 \times n2) \times \cdots \times(md \times nd)}$. Next, the reshaped table is decomposed into subtables. The kth subtable shape is $C(k) = \mathbb{R}^{R_{k-1} \times (mk \times nk) \times R_k}$  ($0 < k < d$), where $R$ being rank, which is subembedding dimension, and $d$ the number of subtables set by the user. The rank values are typically set to 16, 32, and 64 and the total number of subtables is set to 3. TT-Rec results in a 327$\times$ reduction with the three subtable configuration and the rank value of 32 \cite{yin2021tt}.

\textbf{CnR Process. } During inference, weight-sharing algorithms reconstruct the original embedding from their subembeddings. For instance, QR-trick retrieves a subembedding from each subtable using hash functions and performs a selected operation on these subembeddings for reconstruction. This study focuses on element-wise vector multiplication because it generally outperforms other operations across various models \cite{shi2020compositional}. On the other hand, TT-Rec reconstructs a full embedding by fetching \( n1 \), \( n2 \), and \( n3 \) consecutive subembeddings from each subtable and performing mathematical operations on every permutation of these subembeddings. As the entry of the first and third subtables is a vector and that of the second subtable is a matrix consisting of $R$ vectors, the reconstruction process comprises a GEMV operation followed by a dot product for each permutation. 






\section{Motivation} \label{motivation}
\begin{figure*}[!t]
\centering
\includegraphics[width=\textwidth]{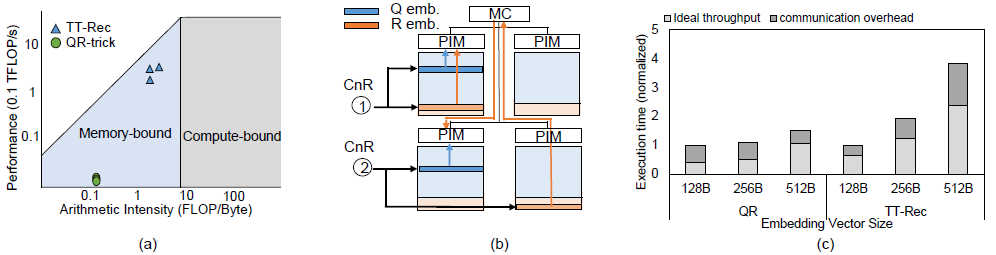}
\caption{Rooline analysis and CPU-PIM communication overview. (a) Roofline analysis. (Evaluated on NVIDIA Tesla V100 with varying batch sizes and total embeddings per GnR.) (b) CPU-PIM communication example (c) Execution time breakdown on weight-sharing algorithms when bankgroup level PIM is utilized. Ideal throughput stands for throughput without CPU-PIM communication.}
\label{fig:communication}
\end{figure*}

\subsection{Weight-Sharing Algorithm Analysis}

\textcolor{black}{Each method in weight-sharing algorithms represents complementary characteristics that provide different options for optimizing the model inference. QR-trick involves a lightweight CnR operation of element-wise multiplication, which amplifies the memory access only twice but degrades the model accuracy. In contrast, TT-Rec results in higher compression and retains the accuracy, but the memory access is largely amplified due to the complicated CnR process. Thus, QR-trick is the solution for the server environment that requires relatively less compression, accuracy, and high throughput, while TT-Rec is a better option for the case where the embedding table size and the accuracy are the primary concerns. Given the importance of both methods, we aim to design a PIM architecture that accelerates both algorithms.}

\subsection{Chances and Challenges of PIM Implementation.} \label{throughput impact}

\textbf{Roofline Analysis. } The roofline analysis \cite{williams2009roofline} in Figure \ref{fig:communication} (a) shows that both algorithms are within the memory-bound region. This indicates that PIM is a promising solution for embedding layers using weight-sharing algorithms, similar to the original embedding layer. The degree of memory-intensive nature differs between the two algorithms. QR-trick is relatively more memory-bounded compared to the TT-Rec. TT-Rec is less memory-bounded than the QR-trick, where its location on the memory-bound region differs for different rank configurations. To accelerate both types of algorithms, we leverage the unique and common characteristics of both algorithms described in the next subsection using the ProactivePIM architecture.

\textbf{CPU-PIM Communication Overhead. } As the CnR process of weight-sharing algorithms consists of vector multiplication or GEMV operation, it poses a distinct challenge to previous NMP/PIM systems for personalized recommendations that only target the GnR process. The CnR process require vector pairs to reside in the same PIM unit for the computation. However, it is impossible to co-locate all vector pairs within the same node because the subtables are distributed across multiple nodes (e.g., ranks and bank groups) for parallel processing. As depicted in Figure \ref{fig:communication} (b), when the vector pair of first CnR lies within the same PIM unit no support from the host is required. However, the second CnR depicts the memory controller required to pass a CnR operand vector that is located in a different node from the other, necessitating CPU-PIM communication. A similar incident occurs for TT-Rec, because its CnR is also composed of vector multiplication. As shown in Figure \ref{fig:communication} (c), the CPU–PIM communication overhead accounts for more than 30\% of the total embedding-layer execution time for both QR-trick and TT-Rec across various embedding dimensions, which is non-negligible in terms of both the throughput and energy efficiency. The drawback is less severe in TT-Rec because the total memory access is mostly attributed to the second subtable, in which a subembedding is a matrix, whereas that of the other subtables is a vector. As subembeddings delivered across bank groups are vectors, the impact of communication is smaller compared to the QR-trick. 

Recently, dynamic techniques have been proposed to alleviate CPU-NMP communications. The inter-node broadcast technique proposed by ABC-DIMM \cite{sun2021abc} alleviates the overhead of sending the same data to each NMP individually, which is particulary effective for graph applications. However, this method faces challenges with weight-sharing algorithms because the embedding layer is implemented in an asynchronous style within the PIM owing to its sparse and irregular data access. The asynchronous nature of PIM operations for the embedding layer requires each PIM to fetch distinct data, resulting in idle cycles as each PIM should wait for its dedicated data to appear on the shared bus. Therefore, the CPU-PIM communication overhead of weight-sharing algorithms coan only be alleviated by point-to-point communication between PIM units. Although interconnect methods for DIMM-NMP \cite{zhou2023dimm} have been proposed to address point-to-point inter-NMP communications, modern PIM architectures still lack support for inter-PIM communication \cite{jonatan2024scalability}. Even when point-to-point communication between bank groups is implemented, channel-to-channel communication between the HBM still exists. Hence, utilizing prior studies can not effectively resolve CPU-PIM communication on weight-sharing algorithms. 

\textbf{Increased Execution Latency. } Even without communication overhead, weight-sharing algorithms incur increased memory access compared to the original embedding layer because CnR requires access to all subtables. This degrades the overall DLRM inference throughput because the embedding layer is a major bottleneck for the inference. Employing a deeper level of PIM to achieve better acceleration could also be a solution; however, it causes extra area overhead in the DRAM die. For instance, bank level PIM incurs roughly 4$\times$ area overhead compared to bankgroup level PIM as four banks comprises a bank group. Coupled with the CPU-PIM communication overhead, another method regarding the algorithm characteristics must be utilized to fully utilize the effective PIM bandwidth.

\subsection{Opportunities for Enhancing PIM Speedup}
\textbf{Concentrated Subembedding Access.} Each subtable must be accessed concurrently to reconstruct the original embedding. This feature results in the total access to embedding not being distributed across subtables but remaining the same for each subtable. In addition, the total number of subembeddings in a subtable is much smaller than that in the original embedding table. Thus, the degree of access to individual subembedding becomes more concentrated compared to the original embedding, which intensifies the access rate. 

\textbf{Intra-GnR Locality.} Another observation is that subembeddings are reused not only across GnRs of the same batch but also within a single GnR. In conventional recommendation systems, a user's interaction with an item is typically reflected once in one sparse feature, thereby preventing multiple accesses within a single GnR. However, as each subembedding in the weight-sharing algorithms is accessed across the CnR of multiple embeddings, subembeddings are reused within a GnR. We term this characteristic the intra-GnR locality. As shown in Figure \ref{fig:dlrm_intro} (b), access to consecutive embeddings cause multiple accesses to a single subembedding within the Q subtable. The intra-GnR locality is further strengthened for smaller subtable sizes, given that access to the subembeddings becomes more concentrated. 

\textbf{Subtable Size Imbalance.} Finally, the size of each subtable differs significantly within each algorithm. For the QR-trick, the R subtable is negligible in size compared with the Q subtable. The R subtable contains a number of entries equal to the hash collision value, whereas the Q subtable's size is derived by dividing the total entries of the original table by the hash collision value, which results in a substantially larger Q subtable. For instance, with a hash collision value of 60, each R subtable is only 0.12MB. For TT-Rec, although the total entries for all tables are roughly the same, the first and third subtable sizes are far smaller than that of the second subtable, as each second subtable entry is a matrix. Again, the size of each 1st subtable is only 0.1MB when assuming 1600 subtable entries with original embedding size of 512B. Previous designs are not optimized to use these characteristics, losing the opportunity for extra speedup.

\section{ProactivePIM Architecture}
We propose ProactivePIM, a PIM architecture that supports weight-sharing algorithms. ProactivePIM employs an integrated approach for speedup by utilizing an in-memory cache (detailed in Section \ref{prefetch}), eliminating communication overhead through data mapping (detailed in Section \ref{mapping_strategy}), and minimizing the additional overhead in the PIM system. \textcolor{black}{The core novelty of our design lies in the architecture (detailed in Section \ref{architecture}) and its synergy with memory access optimizations (detailed in Section \ref{prefetch} to \ref{tt_opt}). ProactivePIM provides the foundation for efficient acceleration, enabling maximal utilization of locality characteristics and reduction of the amplified memory access with a few PIM-HBM stacks. Optimizations such as subtable duplication and maximizing subembedding reuse for TT-Rec are employed for full utilization of the proposed architecture.}

%


\subsection{ProactivePIM Architecture} \label{architecture}

\textcolor{black}{ProactivePIM is integrated into the CPU platform as it offers several advantages over the GPU platform \cite{jain2023optimizing}, \cite{jain2025load}, \cite{jiang2021microrec}. First, DLRM is a memory-bound application, where the off-chip memory accesses in the embedding layer bottleneck the overall performance. Second, batching is limited to small batch sizes to meet the tail-latency requirements, typically fewer than 64. Thus, CPUs have been preferred over GPUs for inference, whereas GPUs have been widely used for training.}

\begin{figure}[htbp]
\centering
\includegraphics[width=\columnwidth]{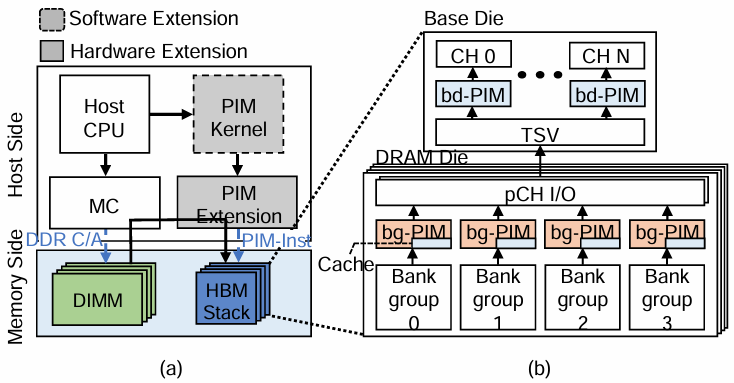}
\caption{ProactivePIM system overview. (a) Host side implementation and heterogeneous memory architecture. MC indicates memory controller. (b) 2-level PIM system overview within HBM2.}
\label{fig:ProactivePIM_overview}
\end{figure}

The overall architecture is depicted in Figure \ref{fig:ProactivePIM_overview} (a). From a top-level view, the ProactivePIM system utilizes a heterogeneous memory architecture comprising multiple HBM stacks and DIMMs. By leveraging the long-tail phenomenon inherent in weight-sharing algorithms, frequently accessed subembeddings are offloaded to the HBM, whereas DIMMs accommodate the remainder, reducing the compression burden on the HBM and enabling scalable model inference. Frequently accessed subembeddings are profiled during training and offloaded to the HBM by the PIM kernel before inference, with the offloading ratio determined by the bandwidth ratio between the PIM-integrated HBM and the DIMMs. Prior studies \cite{kal2021space, ke2020recnmp} have shown that profiling incurs minimal overhead in the end-to-end inference time. Thus, the utilization of heterogeneous memory bandwidth is maximized. 

A two-level PIM system is integrated inside the HBM to leverage parallel processing within the memory. \textcolor{black}{This hierarchy is adopted to support a broad range of current and future DLRM workloads, accommodating the yearly growth in memory intensity. In addition, it can maximally utilize the strong locality of weight-sharing algorithms as an expanded bandwidth ratio between two memory systems necessitates transferring more embeddings to the PIM-HBM side.} As shown in the Figure \ref{fig:ProactivePIM_overview} (b), the first-level processing unit (bd-PIM) is on the base die, whereas the second-level units (bg-PIM) are inside each bank group. Each bg-PIM performs CnR followed by GnR on the CnR outputs. Once all computation are complete, the partial results are passed to the bd-PIM for the final reduction. The SRAM cache is placed in each bg-PIM to utilize the intra-GnR locality of subembeddings, reducing bank access during embedding reconstruction. As in previous works \cite{ke2020recnmp, park2021trim}, the PIM kernel launched by the host CPU sends PIM requests to the memory controller extension (PIM extension), which schedules and issues PIM-Inst to start CnR/GnR. If the subembeddings are in the DIMM, they are read by the memory controller, passed through the PIM extension, and packed into the PIM-Inst. We employ HP method for data placement because its energy efficiency is better than VP.

\subsection{Hot Subembedding Prefetch} \label{prefetch}

\begin{figure}[htbp]
\centering
\includegraphics[width=0.98\columnwidth]{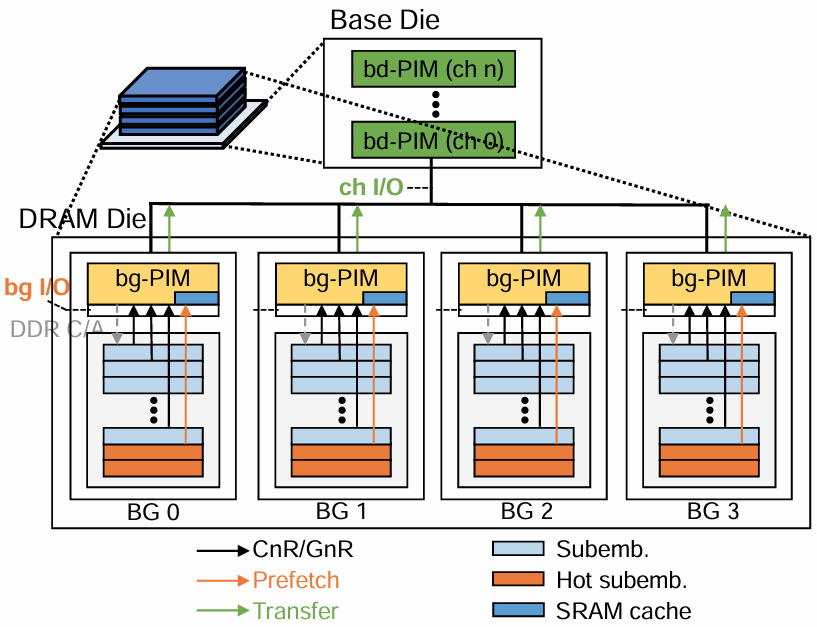}
\caption{Hot subembedding prefetch overview. Prefetch and transfer use a distinct data bus, which enables prefetch to be pipelined with data transfer.}
\label{fig:hot_subemb_overview}
\end{figure}

\textbf{Prefetch Overview. } To leverage intra-GnR locality, we propose a unique prefetch method that offloads frequently accessed "hot" subembeddings into the bg-PIM cache prior to embedding operation execution. Figure \ref{fig:hot_subemb_overview} depicts an overview of hot subembedding prefetch. Before the CnR/GnR operations, hot subembeddings are read from the columns and cached into the SRAM (indicated by orange arrows). During the CnR/GnR phase, most subembeddings are read from the columns (indicated by black arrows), but access to the hot subembeddings is directed to the cache, eliminating the need to access the columns, thus reducing memory access. Finally, the partial sums, which are the result of the CnR/GnR operations, are transferred to the bd-PIM for final reduction (indicated by green arrows).

Our key contribution lies in designing an efficient prefetch method while minimizing the area overhead within both the PIM extension and the bg-PIM cache. Integrating a cache within the PIM requires additional registers because the PIM extension needs to store and check the physical addresses of the cached data to maintain the correct timing for the command issue. In the ProactivePIM system, we mitigate this overhead by prefetching the entire subembedding of a subtable stored in the HBM rather than individual subembeddings, such that the PIM extension only needs to verify the subtable ID when issuing cached data access. Moreover, as described in Section \ref{mapping_strategy}, specific subtables are replicated in each bank group to reduce CPU-PIM communication, which facilitates prefetching of an entire subtable because its start and end addresses are consistent across all PIM units. However, prefetching multiple subtables increases the cache area overhead and prefetch complexity, which necessitates prefetching only a single subtable.

\textbf{Target Subtable to Prefetch.} The key criteria for selecting a target subtable are (i) the degree of intra-GnR locality and (ii) the subtable size, both of which influence the cache efficiency. In the QR-trick, the R subtable is much smaller than the Q subtable, and the subembeddings exhibit high intra-GnR locality as described in Section \ref{weight-sharing-intro}. Therefore, the R subtable is chosen.  For TT-Rec, the intra-GnR locality is similar among all the subtables since they have nearly the same total entries. However, the second subtable is three-dimensional, resulting in a larger size than the others. Furthermore, as described in Section \ref{tt_opt}, the third subtable is read from the bankgroup during the skinny GEMM operation. Therefore, the first subtable is the prefetch target.

\begin{figure}[htbp]
\centering
\includegraphics[width=\columnwidth]{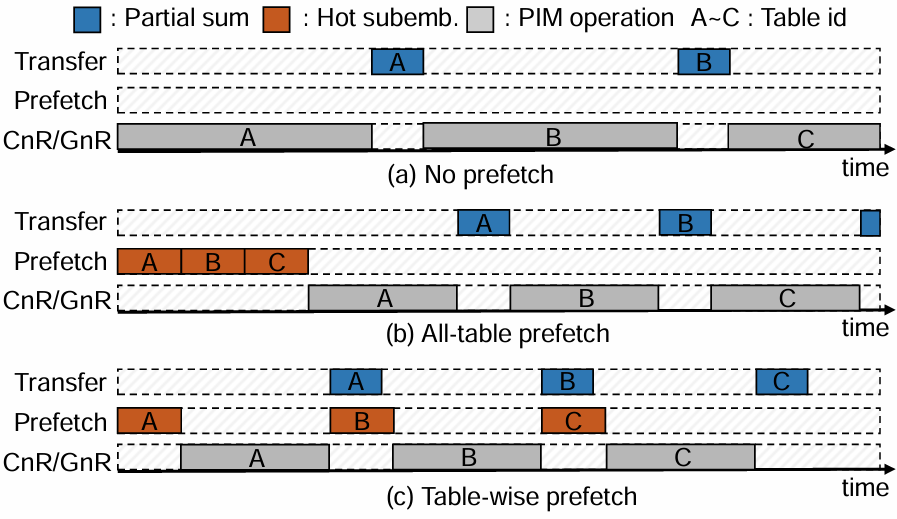}
\caption{CnR/GnR execution timeline without and with prefetch. (a) Original embedding operation (GnR) execution timeline. (b) Reduced execution timeline when all-table prefetch is applied. (c) Execution timeline achieved with table-wise prefetch. }
\label{fig:timeline_comparison}
\end{figure}

\textbf{Table-wise Prefetch.} Prefetching hot subembeddings leads to a notable reduction in the total execution time by lowering memory access. Figure \ref{fig:timeline_comparison} (a) shows the original execution timeline for CnR/GnR execution of the three original tables. Once CnR/GnR completes for table A, the partial sums from each bg-PIM are transferred to the bd-PIM during the transfer phase. Figure \ref{fig:timeline_comparison} (b) illustrates the naive method in which all target subtables are read and offloaded before the embedding layer execution, which we name the all-table prefetch. At the cost of the initial delay, the time consumed in the PIM operation is largely reduced owing to the cache hit, which is permanent before the inference terminates. However, this method requires a large SRAM cache as it needs to store every target subtable of all original tables, which leads to 14\% area overhead in DRAM die for 5-table configuration. This effect worsens with more tables. VP can be utilized to reduce the SRAM size fourfold by distributing subembedding dimensions across bank groups, but the data read per access falls below the DRAM burst granularity for dimensions smaller than 256B, requiring another approach. 


To benefit from locality characteristics, previous PIM architectures adopted table-wise scheduling for the batch process of embedding operations, observing that a batch of requests could be executed in a table-wise fashion (rather than request-wise) within the PIM. Building on this method, ProactivePIM employs a table-wise prefetch scheme that requires only a single subtable to be loaded at a time. This approach offloads a target subtable of the next table once the CnR/GnR of the previous table completes, as illustrated in Figure \ref{fig:hot_subemb_overview}. Before CnR/GnR begins in table B, the target subtable is prefetched to the cache during table A’s data transfer phase using the same process as in tables B and C. This method allows the SRAM cache to accommodate only a single subtable, thereby significantly reducing the PIM area overhead. In addition, the partial sum transfer and prefetching are pipelined using separate buses to minimize the prefetch overhead in the execution time. As shown in Figure \ref{fig:pim_architectures} (a), transfer uses an external bus (channel I/O), whereas prefetching uses an internal bus (bankgroup I/O). Consequently, when the PIM-Inst for data transfer is issued from the decoder, bg-PIM starts to prefetch.

As shown in Equation \ref{eq1}, the transfer latency (\textit{tTransfer}) is proportional to the total number of bank groups in a channel, read command delay between bank groups (\textit{tccd\_s}), batch size, and embedding dimension (\textit{vlen}). In contrast, the prefetch latency (\textit{tPrefetch}) is proportional to the total number of subembeddings within a target subtable (\textit{totsub}), read command delay within bank groups (\textit{tccd\_l}), and subembedding dimension (\textit{subvlen}). Because a small batch size can cause severe load imbalance across PIM nodes, which is especially critical for a multi-stack HBM system, we choose a batch size of 16. As described in Section \ref{eval}, the prefetch overhead is minimal, indicating that \textit{tPrefetch} is similar to \textit{tTransfer}. The host reads the memory-mapped register (MMReg) placed within the PIM to verify prefetch completion and enable the next CnR/GnR execution. The target subtable data are mapped consecutively in a row to utilize the row buffer hit.

\begin{align} \label{eq1}
tTransfer = batch \times vlen& \times bank groups \times tccd\_s \\
tPrefetch = tot sub &\times subvlen \times tccd\_l
\end{align} 

\subsection{Subtable Mapping Strategy} \label{mapping_strategy}

\begin{figure}[htbp]
\centering
\includegraphics[width=0.9\linewidth]{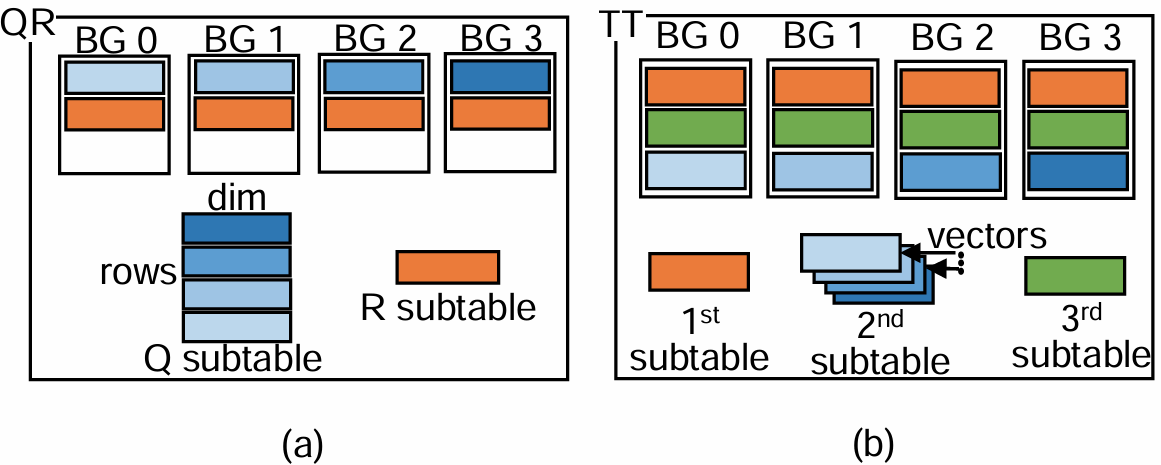}
\caption{Subtable mapping strategy. (a) QR-trick mapping. (b) TT-Rec mapping. 2nd subtable subembeddings are distributed across bank groups for maximizing cache efficiency.}
\label{fig:data_mapping}
\end{figure}

\textbf{Overview.} To eliminate the CPU-PIM communication of weight-sharing algorithms, ProactivePIM devises a subtable mapping strategy. The core idea of this strategy is based on the observation that the number of entries in each subtable differs significantly. ProactivePIM replicates small subtables into every PIM node so that element-wise multiplication is performed without CPU-PIM communication. As shown in Figure \ref{fig:data_mapping} (a) and (b), QR-trick duplicates the R subtable across bank groups, whereas in TT-Rec, the 1st and 3rd subtables are duplicated. The remaining subtables in both methods are distributed among the bank groups. 

\textbf{Duplication Overhead Analysis.} The duplication overhead becomes critical as memory access parallelism deepens, particularly as the number of original tables increases. For instance, with a 512B embedding size, the overhead of each R subtable is 30KB with a hash collision of 60 in QR-trick and 100KB for the 1st and 3rd subtable each when assuming a rank value of 16 and 1600 total subembedding entries in TT-Rec. For the 100 embedding tables, the total storage occupied by each algorithm is 3MB and 10MB. As the bank group capacity of HBM2 is 256MB, the capacity overhead from duplication is less than 3.9\% in HBM2 at the bank group level. However, this overhead increases significantly at the bank level, reaching a maximum of 15.6\%. ProactivePIM uses bank group-level PIM to reduce memory capacity overhead. The duplication overhead for the QR-trick is influenced by hash collision, whereas for TT-Rec, it depends on the rank value and roughly scales with the cube root of the number of original table entries. In both algorithms, the overhead is also affected by the embedding size and the number of tables.

\textbf{Preprocessing. } To maintain the duplication overhead within the predefined threshold and the size of each prefetch target subtable in the HBM below the SRAM cache size (to ensure a correct table-wise prefetch operation), the PIM kernel fine-tunes the subtable duplication policy before inference. When the target subtable exceeds the threshold, the kernel exchanges subembeddings between memories. Subembeddings from the target subtables are freed from duplication and then moved from the HBM to the DIMM, while those from other subtables are simultaneously moved in the opposite direction, without duplication. The exchange is executed by considering the overall access rates to each memory, ensuring that the access ratio matches the bandwidth ratio between the HBM and the DIMM. Subembeddings with the lowest access rates are selected as victims of the HBM, while those with the highest access rates are selected from the DIMM. When subembedding offloaded to the DIMM is required for the PIM computation, it is delivered to the target PIM in the QR-trick and is broadcast to all PIM units for TT-Rec. Although bandwidth ratio aware exchange could result in more subembeddings traveling from DIMM to HBM than in the opposite direction, extra capacity in the HBM is not required because those removed from the HBM are freed from duplication across bank groups. In addition, despite the target subtable of table-wise prefetch being distributed between two memories with this process, the prefetch effect on speedup is not compromised, as subembeddings with low access rates are first offloaded to the DIMM.

\begin{figure}[htbp]
\centering
\includegraphics[width=\columnwidth]{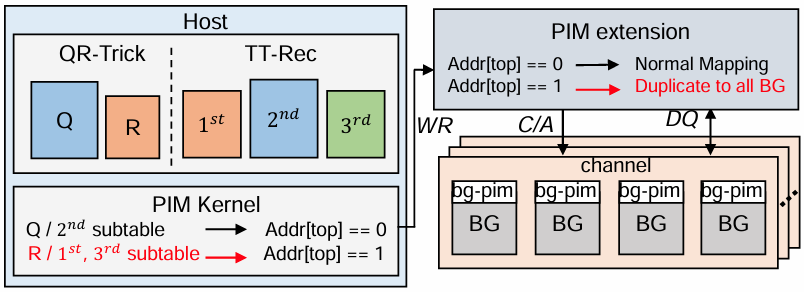}
\caption{Preprocessing scheme of subtable mapping strategy.}
\label{fig:preprocessing}
\end{figure}

After the subtable duplication policy is determined, the PIM kernel delivers requests to the PIM extension that executes the duplication process. Figure \ref{fig:preprocessing} depicts the preprocessing overview. First, the user specifies the subtables to duplicate using the kernel function. Then, the kernel sends write commands with the top bit of the physical address set to 1 for the physical address of the subtable to be duplicated. The PIM extension detects this bit and delivers write commands to all bank groups by altering the bank group bits in the physical address. Once the duplication is complete, the kernel allocates the remaining subtables to the available spaces in the HBM. During inference, the extension checks the top bit and adjusts the channel and bank group in the PIM-Inst for duplicated subtables.

\subsection{TT-Rec Optimization on ProactivePIM} \label{tt_opt}

\textbf{Maxmizing Subembedding Reuse.} The naive implementation of TT-Rec CnR leads to inefficient memory access, as its computation is composed of a permutation of each subtable subembeddings. Assuming that the CnR process requires $n1$, $n2$, $n3$ subembeddings from each subtable, static cache utilization reduces the total computation from $n1 \times n2 \times n3$ to $n2 \times n3$, which is still memory-intensive compared to CPU execution that fetch $n1 + n2 + n3$ subembedding when a full on-chip cache hit occurs. We observe that TT-Rec CnR can be divided into two stages of batched skinny GEMM, where the first stage operands are 1st and 2nd subtable, while the second stage operands are the resulting matrix of the first stage and the 3rd subtable. We aim to reduce duplicate access to subembeddings by maximizing the reuse of the skinny matrix. As each 2nd subtable subembedding required for CnR is determined during runtime, two-stage batched skinny GEMM is processed across PIM units where each PIM runs one or a few skinny GEMM.

\begin{figure}[htbp]
\centering
\includegraphics[width=\columnwidth]{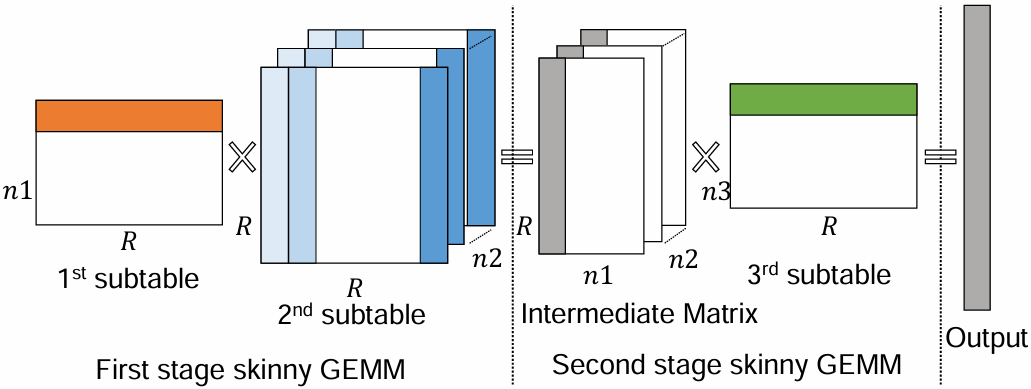}
\caption{Two-stage skinny GEMM computation process of TT-Rec CnR. The n2 dimension of 2nd subtable and intermediate matrix is distributed across bg-PIM units.}
\label{fig:skinny GEMM}
\end{figure}

Figure \ref{fig:skinny GEMM} depicts the corresponding implementation. As shown in the figure, TT-Rec CnR is reordered into a two-stage batched skinny GEMM. The skinny matrix of each stage is stored within static cache to prevent repeated access, which is forwarded to the multiply-and-accumulate (MAC) unit, which is processed by subembedding from another matrix that is read from the bank group. In contrast, the first-stage skinny matrix (which is the 1st subtable subembedding), the skinny matrix of the second stage is not stored within the static cache before CnR begins. Therefore, we reserve an extra buffer for the intermediate matrix to support the full execution. Assuming a 512B original embedding size and a rank value of 32, only a 0.25KB extra overhead is introduced, which is negligible regarding the cache size required for hot subembedding prefetch. With this computation flow, the total memory access is significantly reduced, making it comparable to that of CPU execution.

\subsection{ProactivePIM Implementation}

\begin{figure}[htbp]
\centering
\includegraphics[width=0.98\columnwidth]{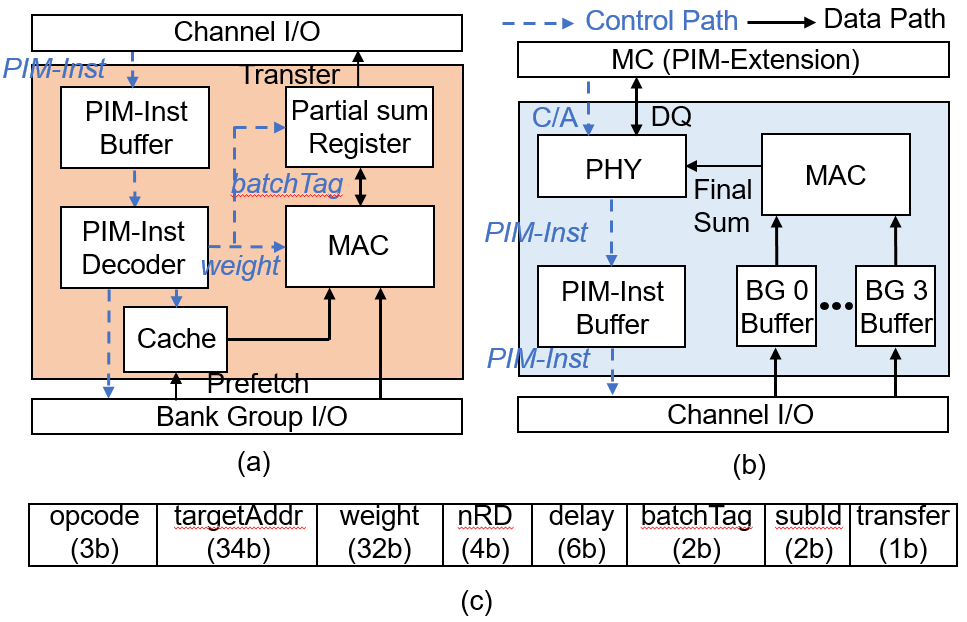}
\caption{ProactivePIM architecture implementation. (a) bg-PIM architecture. (b) bd-PIM architecture. (c) PIM-Instruction format.}
\label{fig:pim_architectures}
\end{figure}

\textbf{PIM Architecture.} Figure \ref{fig:pim_architectures} (a) shows bg-PIM, which is equipped with a PIM instruction buffer, PIM instruction decoder, MAC unit and partial sum register. PIM instructions (PIM-Inst) delivered from the host are buffered and decoded in the decoder to set the control paths of the PIM unit. MMReg is placed within the decoder to indicate the start and end of the prefetch address. To fully support both element-wise multiplication and skinny GEMM operations, we placed 64 MAC units for CnR/GnR. The SRAM cache size is configured as 40KB to match a 1:1 ratio between logic and cache area, leveraging intra-GnR locality while maintaining the PIM area overhead in a low range. The computational capability for skinny GEMM and the integration of SRAM cache enables ProactivePIM to support the efficient inference of a wide range of memory-bound operations \cite{he2025papi, gonzalez2012powergraph, yun2022grande}. Figure \ref{fig:pim_architectures} (b) shows the bd-PIM, which is similar to the bg-PIM without the cache. \textcolor{black}{We assume that the bd-PIM is located near the global I/O interface, and each bg-PIM is placed next to the bank-group I/O periphery logic, without requiring modification of the existing DRAM interface and TSV.}

\textbf{PIM Instruction.} Figure \ref{fig:pim_architectures} (c) depicts the PIM-Inst format. The \textit{opcode} (3-bit) specifies the operation type that the bg-PIM should execute, the \textit{target address} (34-bit) denotes the start address of the data to be read from the node, and the \textit{weight} (32-bit) delivers the weight value for the weighted sum. Note that the opcode is extended from previous works, including the \textit{load to the input register}, and \textit{store the result in the SRAM cache}. The \textit{nRD} (4-bit) indicates the subembedding size, and the \textit{delay} (6-bit) indicates the time for instruction decode after being delivered to the bg-PIM. The \textit{batchTag} (2-bit) is the assigned batch of the instruction, and the \textit{subid} (2-bit) indicates the subtable ID, used for CnR termination. Finally, \textit{transfer} (1-bit) determines whether the partial sum is complete and ready to deliver. 

\textbf{Memory Model.}  We use a direct mapping method to support PIM operations and allocate memory space for ProactivePIM as uncachable to avoid cache coherence. To utilize the heterogeneous memory systems, ProactivePIM makes use of a flat addressable method, which has been continuously employed in prior studies \cite{kal2021space, chou2014cameo, sim2014transparent}. Additionally, to handle the bandwidth limit of command/address (C/A) pin, we follow the prior method \cite{park2021trim}, which utilizes DQ pins along with C/A pins to deliver PIM-Inst to bg-PIMs within a short cycle. Specifically, we utilize 14 C/A pins with 128-bit wide DQ pins to transfer one PIM instruction in a single cycle. Owing to the abundant HBM DQ pins, ProactivePIM successfully addresses the C/A bandwidth overhead.

\section{Experimental Setup}
\textbf{Simulation Methodology. } We implemented ProactivePIM in DRAMsim3 \cite{li2020dramsim3}, a cycle-accurate memory simulator. HBM2 $\times$128 and DDR4-3200 $\times$8 were set as the base memory systems, and Table \ref{tab:system_parameters} summarizes the specifications. We compare ProactivePIM performance against RecNMP, TRiM and SPACE. Hot subembeddings are treated as hot embeddings to utilize existing designs that exploit temporal locality. We generated synthetic traces of CnR/GnR from the official DLRM code using the criteo datasets \cite{criteo_kaggle_ad_dataset, reddi2020mlperf} that are available online. \textcolor{black}{In addition, we generate traces with MIX1 dataset introduced in SPACE, composed of Amazon datasets \cite{ni2019justifying} and Google Maps \cite{pasricha2018translation}.} The embedding lookup per GnR was set to 80, with dimensions ranging from 128B to 512B. To ensure a fair comparison, we treat all subembeddings of the target subtable as hot vectors, allowing for the maximal exploitation of techniques that utilize temporal locality, as proposed in prior works. For instance, the R subtable has the highest priority when the reduction locality technique is employed, storing the reconstructed embedding of hot subembeddings in the available space in HBM2. For RecNMP, we linearly scaled the speedup of RankCache, following TRiM.



\newcolumntype{M}[1]{>{\centering\arraybackslash}m{#1}}

\begin{table}[htbp]
\normalsize
\centering
\caption{Memory System Specification}

\label{tab:system_parameters}

\resizebox{\columnwidth}{!}{
\begin{tabular}{M{0.25\columnwidth}M{0.75\columnwidth}}
\Xhline{2\arrayrulewidth}
\multicolumn{2}{c}{HBM2 $\times$128 device} \\
\Xhline{2\arrayrulewidth}
Component &  1 stack of HBM2 (4GB)\\
\hline
Memory Organization & 8 channels per 4-hi stack, 4 bank-groups per channel, 4 banks per bank-group \\
\hline
Timing Parameters & tCL=14,  tRP=14,  tRCD=14,  tCCD\_S=1,  tCCD\_L=2,  tBL=4 \\ 
\hline
Clock Frequency & 1000MHz \\
\Xhline{2\arrayrulewidth}

\end{tabular}
}

\vspace{8pt}

\resizebox{\columnwidth}{!}{
\begin{tabular}{M{0.25\columnwidth}M{0.75\columnwidth}}
\Xhline{2\arrayrulewidth}
\multicolumn{2}{c}{DDR4-3200 $\times$8 device} \\ 
\Xhline{2\arrayrulewidth}
Component &  2 DIMM module (16GB)\\
\hline
Memory Organization & 1 channel per DIMM, 2 ranks per channel, 4 bank-groups per rank, 4 banks per bank-group\\
\hline
Timing Parameter & tCL=22,  tRP=22,  tRCD=22,  tCCD\_S=4,  tCCD\_L=8,  tBL=8 \\
\hline
Clock Frequency & 1600MHz \\
\Xhline{2\arrayrulewidth}

\end{tabular}
}
\end{table}

\textbf{Area Overhead and Power Consumption. } Bd-PIM and bg-PIM were synthesized using a Synopsys Design Compiler with 45nm CMOS technology operating at a 300MHz clock frequency to obtain the area overhead and power consumption. The area was scaled to a 20nm DRAM process, taking into account that the DRAM process is 10$\times$ less dense than the ASIC process \cite{devaux2019true, shin2018mcdram, park2021trim}. \textcolor{black}{The SRAM cache power and area were estimated using the CACTI 6.5 \cite{muralimanohar2009cacti} with 32nm process, where the area was scaled following the same process for bg-PIM. It is worth noting that the assumption of applying a 10$\times$ density penalty when mapping SRAM logic into the DRAM process is approximately 5$\times$ more conservative compared to the previous work \cite{gu2020ipim}.}  The power consumption of the memory and off-chip I/O is calculated following \cite{o2017fine}.



\section{Evaluation Result}
\begin{figure*}[!t]
\centering
\includegraphics[width=\linewidth]{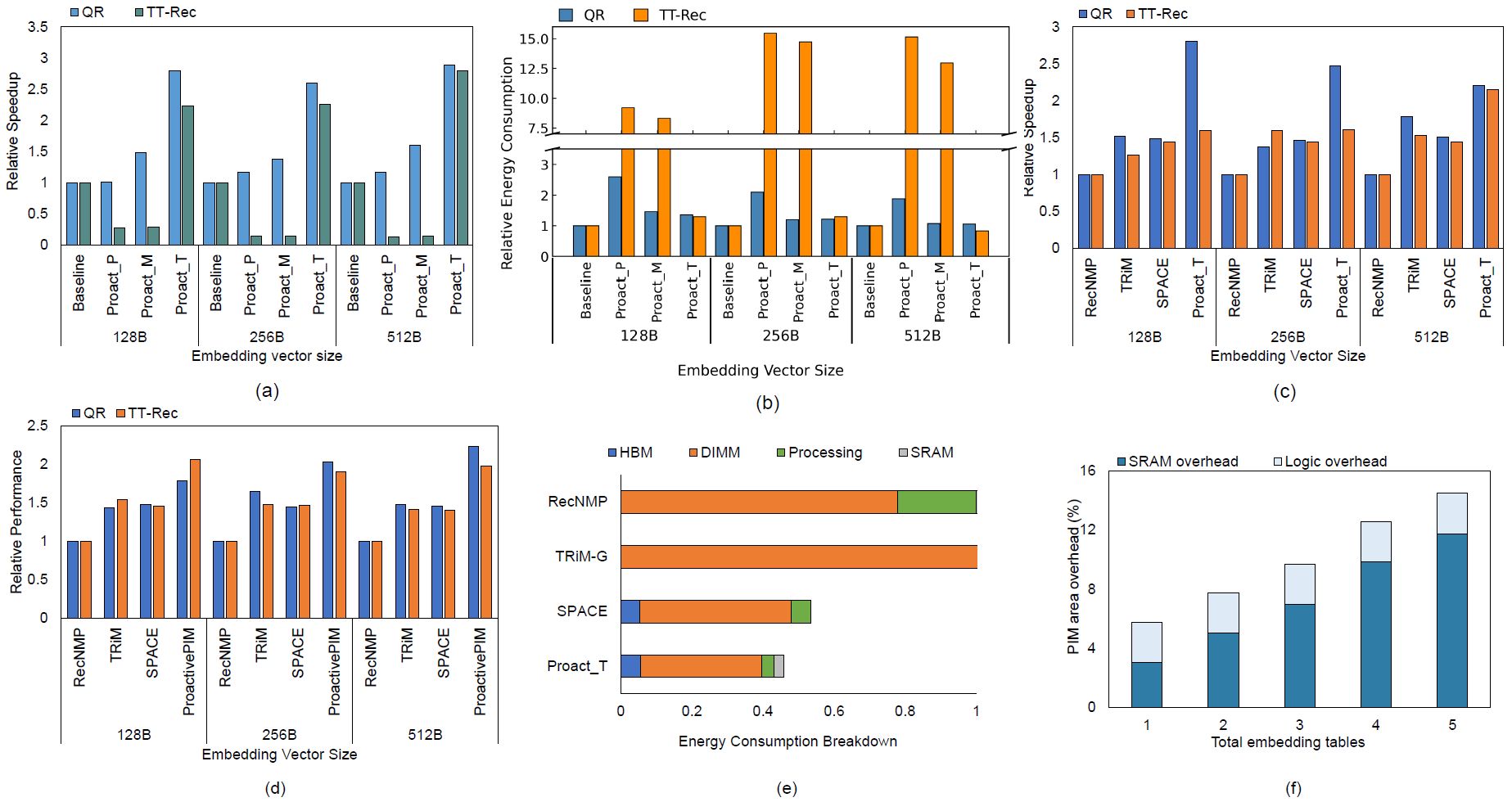}
\caption{Performance of design choices on QR-trick/TT-Rec. (a) Relative speedup when design choices are gradually incorporated. (b) Relative energy consumption for design choices. (c) Relative speedup compared to the original embedding layer on normal PIM and ProactivePIM systems for \textcolor{black}{the Criteo dataset}. Results are normalized to the baseline. (d) \textcolor{black}{Relative speedup compared to previous PIM and ProactivePIM systems for MIX1 dataset.} (e) Energy consumption breakdown for QR-trick. (f) Cache sensitivity study}
\label{fig:design_optimization_perf_impact}
\end{figure*}

We evaluate the impact of ProactivePIM’s design choices as the first step in analyzing the effectiveness of each choice. We then compare the performance of ProactivePIM with state-of-the-art NMP designs, which target the embedding operation. In addition, the area overhead and power consumption introduced to the overall system for the ProactivePIM architecture support are analyzed. 

\subsection{Performance Impact of Design Optimizations}

\textcolor{black}{We first evaluate the impact of our design choices on the speedup on the Criteo dataset.} Figures \ref{fig:design_optimization_perf_impact} (a) and (b) shows performance improvements in speedup and energy efficiency improvements when design optimizations are gradually incorporated within the ProactivePIM system for the QR-trick and TT-Rec. The baseline is normal HBM2 with the last-level cache of the host CPU set to 32MB. The first two configurations, Proact\_P and Proact\_M, refer to PIM systems without SRAM cache: Proact\_P adds only bg-PIM to the baseline, whereas Proact\_M applies the subtable mapping strategy to Proact\_P. Finally, Proact\_T represents our complete PIM system with table-wise prefetch and TT-Rec optimization. 

With a 512B embedding size, Proact\_P and Proact\_M achieve speedups of a 16\% and 60\% for the QR-trick, while 87\% and 86\% degradation occurs in TT-Rec. The performance of Proact\_P is lower than the ideal bankgroup-level PIM speedup, which is $2\times$ (total bank groups $\times$ tccd\_s/tccd\_l) in HBM2. The main cause of the nonideal speedup in Proact\_P is attributed to the CPU-PIM communication overhead for data movement. In contrast, Proact\_M achieves performance improvement than Proact\_P by eliminating the overhead. However, due to the amplified memory access of TT-Rec, both configurations show poor performance. In each algorithm, Proact\_T achieves $2.89\times$ and $2.78\times$ better results than the baseline due to the utilization of intra-pooling locality with the table-wise prefetch support on SRAM cache. In addition, the TT-rec optimization technique removes amplified memory access, fully utilizing the PIM capability. Although prefetch overhead exists when \textit{tTransfer} is less than \textit{tPrefetch}, the overhead is minimal because the speedup of Proact\_T over Proact\_M that is achieved by the cache is nearly ideal.

\textcolor{black}{We perform an experiment to determine how close the mapping strategy effect is to the ideal throughput in Section \ref{motivation}. For 128B, 256B, and 512B embedding sizes, subtable mapping achieves 0.06\%, 0.91\%, and 3.4\% higher throughput in QR-trick, and 0.63\%, 0.23\%, and 0.06\% lower throughput in TT-Rec. Generally, the mapping strategy mitigates load imbalance caused by the duplication of target subtables while introducing bank conflicts for every bank group. In QR-trick, load balancing is effective because the extremely small R subtable is located in a few bank groups. In TT-Rec, bank conflict becomes more significant because the load imbalance from the mapping target subtables is negligible, as the 2nd subtable subembedding is substantially larger. Throughput is less reduced for larger embeddings because the increased number of subembedding reads in the CnR process amplifies load imbalance, which is alleviated by the mapping strategy.}

As shown in Figure \ref{fig:design_optimization_perf_impact} (b), the increased memory access of the weight-sharing algorithms energy consumption in the Proact\_P and Proact\_M configurations. Assuming 512B embedding size, QR-trick introduces $1.88\times$ and $1.07\times$ more energy cosnumption, whereas TT-Rec consumes $15.14\times$ and $12.95\times$ amplified energy for each configuration. Proact\_T produces 6\% overhead and 17\% energy savings compared to the baseline, where the baseline shows better energy efficiency for certain configurations owing to the on-chip cache hit regarding the power-law distribution of embedding access. Compared to other PIM configurations, Proact\_T outperforms the other PIM configurations in terms of energy efficiency owing to decreased row activations and read energy.

\subsection{Performance Evaluation with Previous NMP Designs}

To demonstrate the effectiveness of our proposal, we compare the throughput and energy consumption performance of ProactivePIM over prior PIM systems, RecNMP, TRiM-G, and SPACE. For a fair comparision, all configurations utilize HBM2 as the base memory architecture and employ all optimization techniques without a table-wise prefetch, as it requires an extra SRAM cache within the NMP/PIM architecture. TensorDIMM is excluded in this experiment as its configuration in HBM2 results in inefficient memory access below the memory access granularity. Figure \ref{fig:design_optimization_perf_impact} (c) and (d) shows the relative speedup of each previous work normalized to RecNMP for each algorithm, where embedding size ranges from 128B to 512B. When using criteo dataset, Proact\_T achieves $2.22\times$, $1.23\times$, $1.45\times$ speedup over RecNMP, TRiM, and SPACE for QR-trick and improvements of $2.15\times$, $1.41\times$, $1.49\times$ for TT-Rec under the same configurations for an embedding size of 512B. \textcolor{black}{The results are similar for MIX1 dataset, where the speedup is $2\times$, $1.4\times$, $1.41\times$ for QR-trick and $1.76\times$, $1.51\times$, $1.52\times$ for TT-Rec.} The superior performance of ProactivePIM results from efficient cache utilization by leveraging intra-GnR locality. The impact becomes more significant for large dimensions owing to the increased number of data reads per bg-PIM.

Figure \ref{fig:design_optimization_perf_impact} (e) depicts the relative energy consumption of prior works and ProactivePIM, which are normalized to the result of RecNMP. To properly measure the energy efficiency of the HBM-DIMM architecture over DIMM, two DIMMs are utilized across the systems. Energy consumption is reduced by 54.1\%, 53\%, and 52\% compared to RecNMP, TRiM-G and SPACE when the mapping strategy is applied. The energy efficiency comes from the decreased access owing to the SRAM cache, along with the reduction in static energy from a shorter inference time \textcolor{black}{and the power efficiency of HBM2 over DDR4 \cite{tran2016era}}.

\subsection{Sensitivity Study of Cache Capacity} \label{sensitivity}
We conduct a sensitivity study on cache capacity to prove the necessity of the table-wise prefetch method for SRAM size reduction and to show that SRAM overhead is minimal in this method. As the all-table prefetch stores all the target subtables, the area overhead increases as more embedding tables are employed. Figure \ref{fig:design_optimization_perf_impact} (f) shows the variation in PIM area overhead when the total embedding table increases. When the number of tables is 5, the area overhead of the PIM on the HBM2 die is approximately 21\%.

\subsection{Design Overhead}

\textbf{Hardware Overhead.} The total area occupied by the bg-PIM is 5.52$mm^2$ per HBM2 die, which accounts for 5.75\% of each HBM2 die. The SRAM cache accounted for 53\% of the total PIM area. Each PIM includes 64 MAC units and register files for buffer and logic support. The SRAM cache size is set to 40KB to match the logic and cache area ratio of 1:1. The total gate count is 322,578 per bg-PIM. The energy consumption of the SRAM cache accounts for 0.04\% of the total energy consumed by ProactivePIM, which enables high energy savings from DRAM access. Bd-PIM area is 0.33$mm^2$ per channel, with a gate count of 166,697 each. \textcolor{black}{Finally, the PIM extension area overhead is 0.004$mm^2$, with total gate count of 5,642.}

\textbf{Software Overhead.} \textcolor{black}{We evaluated the HBM-DIMM preprocessing runtime on an Intel Xeon Silver 4214 CPU-based server, where the result was 0.01s when using the Criteo Terabyte dataset. In addition, the execution time for generating the hot embedding profile was 1.14 s, similar to the 0.99 s reported in previous work \cite{kal2021space}. This demonstrates that the software overhead of the preprocessing stage is minimal, considering that model update latency can extend to several hours [5]. We also evaluated the execution time for PIM command processing, which is 0.0001s.}

\begin{figure}[htbp]
\centering
\includegraphics[width=0.815\linewidth]{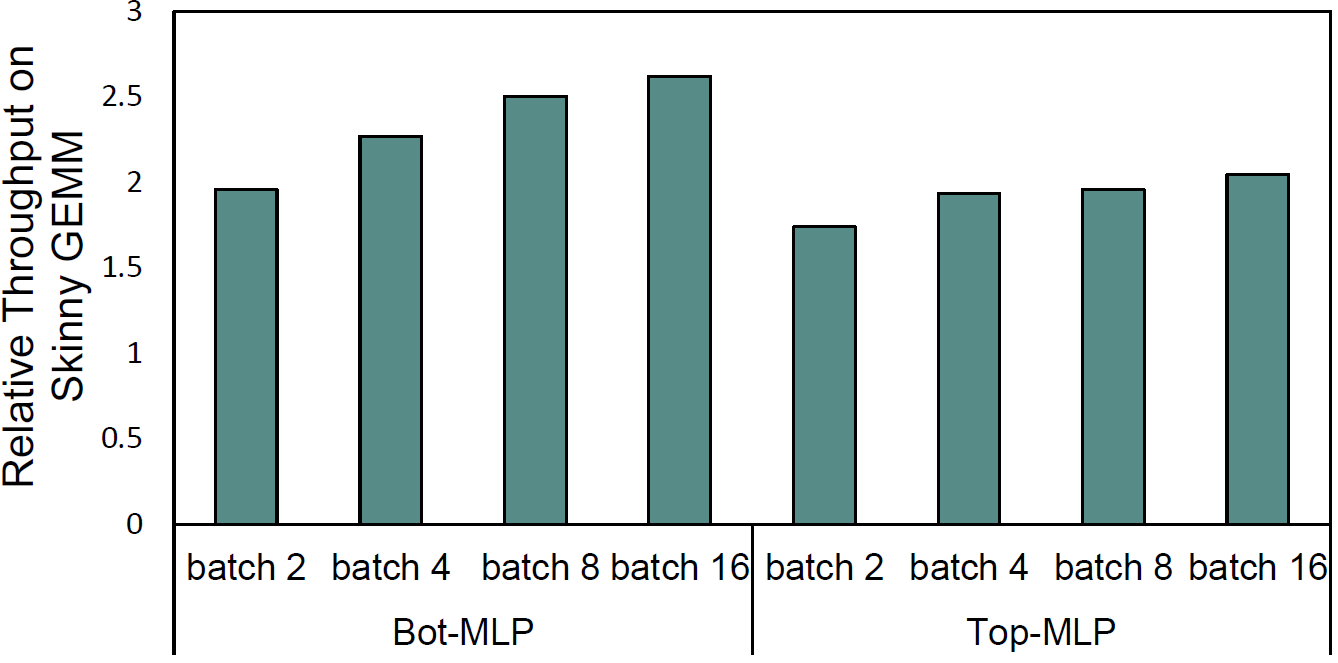}
\caption{ProactivePIM performance on Skinny GEMM Matrix. We use Bot-MLP and Top-MLP of DLRM, with the MLP dimensions of 2560-512-32 and 512-128-1. Result is normalized to normal PIM that can support GEMV operations, such as TRiM and AttAcc \cite{park2024attacc}.}
\label{fig:sensitivity}
\end{figure}

\subsection{ProactivePIM Performance on Skinny GEMM}

\textcolor{black}{We run an experiment on a skinny GEMM workload by varying the batch size from 2 to 16, where bot-MLP and top-MLP were used. The gain was normalized to a normal PIM-HBM baseline that supports GEMV without an SRAM cache, similar to the TRiM and AttAcc \cite{park2024attacc} deployed on the HBM. To fully utilize the parallel processing capability of bg-PIM, we partitioned matrices across bank groups. As shown in the figure below, ProactivePIM achieves more than a 1.7× speedup over the TRiM configuration across all GEMM configurations and batch sizes.  Inference gain of the Top-MLP is less than Bot-MLP as its posterior MLP layer with 128-1 dimensions is not large enough to fully utilize the parallel processing of the bg-PIM. The improvement mainly comes from the reduction of redundant memory accesses by storing the input batch matrix and intermediate results within the SRAM cache and ProactivePIM’s computational capability.}

 \label{eval}

\section{Discussions}
\textbf{Additional Method to Reduce SRAM Capacity.} To further reduce the SRAM capacity within bg-PIM, VP could be an additional method. When using VP, all subembeddings are distributed across the bank groups within a channel. Because each HBM channel contains four bank groups, this approach allows for a fourfold reduction in SRAM size. In the QR-trick, the size of each subembedding is identical to that of the original embedding. However, applying VP to TT-Rec results in a trade-off in bandwidth efficiency. If the rank value is 16 or 32, the subembedding sizes for the 1st and 3rd subtables are 64B and 128B, respectively. With VP, these subembeddings are divided across bank groups, resulting in 16B or 32B per bank group—falling short of the DRAM granularity. This issue is resolved when the rank value is
64, as the subembedding size increases to 256B, which can be evenly distributed as 64B per bank group. However, this configuration reduce the compression rate. Therefore, HP was applied instead of VP to ensure compatibility with both algorithms in this study. 

\textbf{ProactivePIM for Other Applications.} The design principles of ProactivePIM, which exploit data locality patterns with SRAM cache and two-level PIM systems, are not limited to weight-sharing embedding algorithms. \textcolor{black}{ProactivePIM is designed to support GnR and skinny GEMM operations of weight-sharing algorithms. Thus, our design demonstrates generality across a range of memory-bound operations such as GnR in the original DLRM, small batch FC layer \cite{cho2021accelerating}, graph analytic workloads such as pageRank \cite{gonzalez2012powergraph}, and attention in a large-language model. In particular, the integrated SRAM cache can function as a scratchpad to support a small batch FC layer, and it can also accelerate workloads that show long-tail access distribution, such as PageRank with skewed vertex access and the original DLRM. In these cases, profiling must be performed before inference to identify data with high temporal locality to store them within the SRAM cache. During inference, the PIM extension must store locality information so that the access timing is managed within the controller.}

\textbf{ProactivePIM on GPU System. } \textcolor{black}{Although a CPU-only configuration is enough to maximize the GnR/CnR process of the weight-sharing algorithm with ProactivePIM, there exist certain types of recommendation models that employ large MLPs, transforming the model into compute-bound \cite{ke2022hercules, gupta2020architectural}. For those models, our design can be incorporated into a GPU system so that both the MLP and the embedding layer are accelerated. In this case, bandwidth ratio-aware offloading needs to be recalculated as subembeddings within DIMM have to be delivered to the HBM through the PCIe bus.}

\section{Related Works}

\textbf{Processing-In-Memory.} A significant corpus of earlier studies focused on alleviating the memory bandwidth limitations of recommendation systems with NMP architectures \cite{asgari2021fafnir, kal2021space, ke2020recnmp, kwon2019tensordimm, liu2023accelerating, park2021trim}. These studies leveraged the embedding locality characteristics to optimize the embedding layer performance. \textcolor{black}{Compared to prior works, ProactivePIM uniquely targets weight-sharing algorithms compared to prior works such as RecNMP, TRiM, and SPACE by employing SRAM cache within PIM to reduce overall memory access through intensive locality exploitation and by eliminating CPU–PIM communication via subtable mapping.} In recent years, commercial products of PIM have emerged \cite{gomez2021benchmarking, he2020newton, ke2021near, lee2021hardware}. Generally, PIM products adopt bank and bank-group-level in-memory parallelism. Other NMP and PIM architectures target graph processing \cite{lenjani2022gearbox, sun2021abc, tian2023abndp, yun2022grande, zhou2023dimm} and large language models \cite{park2024attacc, seo2024ianus, zhou2022transpim}.

\textbf{PIM Architectures Using SRAM.} ProactivePIM uses 100 KB of SRAM for every 256 MB bank group. UPMEM \cite{devaux2019true} is another example of a PIM system that uses SRAM. In UPMEM, each Data Processing Unit (DPU) has two types of SRAM: IRAM and WRAM, which help support the PIM operations. IRAM and WRAM add up to 88 KB, and each DPU has 64 MB of the main memory. This implies that 88 KB of SRAM is available for every 64 MB of memory. In RecNMP \cite{ke2020recnmp}, 128 KB of SRAM is used for each rank. Stepstone PIM \cite{cho2021accelerating} employs an 8KB scratchpad per DRAM device to accelerate the skinny GEMM inference.

\textbf{Algorithms for Embedding Table Reduction.} Recent studies on embedding table reduction can be classified into three approaches: weight-sharing, mixed-dimension, and low-precision \cite{li2024embedding}. Weight-sharing maps multiple embeddings to a single subembedding, reducing the total entries with some accuracy trade-offs \cite{weinberger2009feature, shi2020compositional, yin2021tt}. Mixed-dimension reduces redundancy by adjusting the dimensions of each entry \cite{ginart2021mixed, lyu2022optembed, luofiited, zhaok2021autoemb}. However, it faces address mapping challenges when applying PIM; using the largest embedding dimension as the default interleaving granularity is necessary to prevent data from being split across multiple PIM units, which in turn leads to wasted memory space for smaller dimensions. Low-precision methods apply binarization or quantization to minimize storage \cite{guan2019post, li2023adaptive}. Nevertheless, this method results in notable degradation in accuracy.

\section{Conclusion}
In this study, we propose ProactivePIM, a two-level PIM architecture designed to accelerate weight-sharing algorithms by exploiting intra-GnR locality and leveraging a subtable mapping strategy. We identify drawbacks of weight-sharing algorithms, such as CPU–PIM communication overhead and increased memory access, while also uncovering unique characteristics such as intra-GnR locality. By utilizing a two-level PIM architecture with an SRAM cache inside the bg-PIM and implementing an efficient prefetch strategy with subtable mapping support, ProactivePIM eliminates CPU-PIM communication overhead and improves the throughput. This results in a $2.22\times$ and $2.15\times$ speedup in QR-trick and TT-Rec, respectively, compared to previous architectures. \textcolor{black}{By supporting model compression algorithms, our method achieves scalability to large-scale recommendation models, which could also be further extended via selective caching of uncompressed embeddings and PIM-enabled CXL support.}


\bibliographystyle{IEEEtranS}
\bibliography{references}

\end{document}